\def\plotone#1{\centering \leavevmode
\epsfxsize=\textwidth \epsfbox{#1}}

\documentstyle[11pt, aaspp4]{article}    

\def\halpha{H$\alpha$}

\def\sqig{$\sim$}

\def\degrees{$^{\circ}$}

\def\source{X0726$-$260}
\def\src{X0726$-$260}
\def\2058{GROJ2058$+$42}
\def\halpha{H$\alpha$}

\lefthead{Corbet et al.}
\righthead{RXTE Observations of \src}

\begin{document}

\title{ RXTE Observations of the Be star X-ray Transient \src\ (4U0728-25) -
Orbital and Pulse Periods}
\author{
Robin H. D. Corbet\altaffilmark{1}
\and
Andrew G. Peele\altaffilmark{2}
}
\affil{Code 662, NASA/Goddard Space Flight Center, Greenbelt, MD 20771}

\altaffiltext{1}{Universities Space Research Association; corbet@lheamail.gsfc.nasa.gov}
\altaffiltext{2}{National Research Council; peele@lheamail.gsfc.nasa.gov}

\begin{abstract}
Rossi X-ray Timing Explorer (RXTE) All Sky Monitor observations of the transient Be star X-ray source
\src\ suggest a 34.5 day period. This is apparently confirmed by a
serendipitous RXTE Proportional Counter Array (PCA) slew detection of the
source on 1997 May 5, near the time of a predicted flux maximum. A
subsequent 5ks pointed observation of \src\ with the RXTE PCA detector
was carried out on 1997 June 7, when \src\ was predicted to be bright again,
and this revealed pulsations at a period of 103.2 seconds. If the 34.5
day period is orbital, then the pulse period is surprisingly long
compared to that predicted by the correlation between orbital period
and spin period observed for
other Be/neutron star systems.  A possible similarity with
\2058\ is briefly discussed.

\end{abstract}

\section{Introduction}
The transient X-ray source \src\ (=4U0728-25, 3A0726-260) has a Be star
optical counterpart that was optically identified by Steiner et al.
(1984). This star was classified as B0Ve by Corbet \& Mason (1984) with
a corresponding distance of 4.6$\pm$1.3 kpc. Negueruela et al. (1996),
however, derive a slightly earlier spectral type of O8-9Ve with a
distance of 6.1$\pm$0.3 kpc.  The X-ray source itself has not been
extensively observed, although \src\ was also detected with Uhuru
(Forman et al.  1978), HEAO-1 (Wood et al. 1984) and ROSAT (Haberl,
cited in Negueruela et al 1996), the only published X-ray light curve
comes from Steiner et al. (1984) who present a rather sparse long term
light curve from the Ariel V Sky Survey Instrument (SSI). Pointed
EXOSAT observations on 1984 Oct. 29 and 1984 Nov. 29 failed to detect
any X-ray emission, even though simultaneous optical observations from
the Isaac Newton Telescope showed the presence of \halpha\ emission
indicating that a circumstellar envelope was still present.

In this paper we present the results of observations of \src\ made with
two instruments on board the Rossi X-ray Timing Explorer (RXTE): the
All Sky Monitor (ASM) and the Proportional Counter Array (PCA). These
observations provide information on the orbital period and spin period
of the neutron star in the system.

\section{Observations}
\subsection{All Sky Monitor}
The ASM detector on board RXTE (Bradt, Rothschild, \& Swank 1983) is
described by Levine et al. (1996). The ASM consists of three similar
Scanning Shadow Cameras, sensitive to X-rays in an energy band of
approximately 2-12 keV, which perform sets of 90 second pointed
observations (``dwells'') so as to cover \sqig80\% of the sky every
\sqig90 minutes.  The analysis presented here makes use of both ``daily
averaged" light curves and light curves from individual dwell data.
The light curve for \src\ in two day bins (derived from the daily
averaged light curve) is shown in Figure 1 and this covers a period of
approximately 1.5 years.  Note that, for comparison, the Crab produces
a mean count rate of 75 counts/s in the ASM.  The mean count rate of
\src\ is 0.14 counts/s or \sqig1.9 mCrab.  However, ASM observations
of blank field regions away from the Galactic center suggest
that background
subtraction may yield a systematic uncertainty of about 0.1 counts/s
(A. Levine, private communication, Remillard 1997).

\subsection{Proportional Counter Array}

The RXTE Proportional Counter Array (PCA )is described by Jahoda et
al.  (1996).  The PCA consists of five individual, nearly identical,
Proportional Counter Units (PCUs) sensitive to X-rays with energies
between 2 - 60 keV with a total effective area of \sqig6500cm$^2$. The
PCUs each have a multi-anode xenon filled volume, with a front propane
volume which is primarily used for background rejection.  For the
entire PCA across the complete energy band the Crab produces a count
rate of 13,000 counts/s. The high-energy HEXTE instrument is coaligned
with the PCA but the results from this instrument are not considered
here due to the faintness of the source and HEXTE's smaller effective
area.

The PCA serendipitously scanned across the position of \src\ on 1997
May 5 (MJD 50573) at a rate of 6 degrees/minute with a minimum distance
of the center of the field of view from the source of approximately
0.2\degrees. The source was clearly detected at a background subtracted
peak rate of 105 counts/s which yields a collimator-response corrected
flux of approximately 10 mCrab (also reported in Corbet \& Peele
1997).  Because of this serendipitous source detection, and a possible
periodicity in the ASM light curve (see Section 3), we undertook a
pointed observation at a subsequent time when the source was predicted
to be bright.

A pointed RXTE observation of \src\ was undertaken on 1997 June 7 (MJD
50606) from 02:54 to 04:54 with an on-source observing time of 5 ks.
Data collection was obtained with a ``Good Xenon'' mode which records
the time and energy of each detected photon with maximum resolution.
During the observation, one of the five PCUs experienced a breakdown
and, to simplify data analysis, we therefore only use data from the
other four detectors.  In addition, we primarily only use data from
xenon layer 1; layer 1 gives better signal to noise than the other
layers but with a reduced high energy efficiency. A comparison of layer
1 data only with data extracted from all three layers in fact reveals
little difference for this observation.  Our analysis presented here
makes use of processed ``real-time'' data which is available more
rapidly than the standard ``level zero'' data products.  The light
curve from this observation is shown in Figure 2.  The source was
clearly detected and also showed significant variability during the
observation.

\section{Results}

A Fourier Transform of the ASM data shows the strongest peak at a
period of approximately 35 days (Figure 3). In addition, the second
strongest peak is at a harmonic of this at \sqig17.5 days.  Using the
Lomb-Scargle method (Lomb 1976, Scargle, 1982) the nominal false alarm
probability (FAP) of the 35 day period is approximately
2$\times$10$^{-3}$ although this does not take into account, for
example, any systematic effects.  A fit of two sine waves (harmonic
and fundamental) to the ASM light
curve yields a period of 34.46 $\pm$ 0.12 days with epoch of maximum
flux at MJD 50365.5 $\pm$ 0.8. 
The serendipitous scan detection of \src\ thus occurred at
a phase of \sqig0.02 and the pointed observation took place at
a phase of \sqig0.97. The light curve folded on this period is
shown in Figure 4 and this shows that, on average, the
source brightens to about 0.4 counts/s (\sqig 5mCrab) for a few days
and is close to 0 counts/s for a substantial fraction of the time.

The Fourier Transform of the PCA light curve (Figure 5) shows a maximum
at a period of \sqig103 seconds with an associated FAP of
$<$10$^{-6}$. A sine wave fit yields a period of 103.2 $\pm$ 0.1 s.
The pulse profile as a function of energy is shown in Figure 6. There
is evidence for a change in profile with energy with a secondary peak
0.5 out of phase from the main peak appearing at higher energies.

The ``classic'' accreting X-ray pulsar model of an absorbed power law
(e.g.  White, Swank, \& Holt 1983) provides an acceptable fit to the
observed 2-20 keV spectrum. While an iron line at \sqig 6.6 keV with an
equivalent width of \sqig 150 eV apparently improves the fit, the
statistical significance of this is difficult to estimate due to the
effects of the Xe L edge in the response matrix. To allow for errors in
the response matrix in a simple way, additional 2\% errors were
included together with the statistical errors when doing the fit.  When
these systematic errors are included, a satisfactory fit is obtained
without the need to include an iron line, and we derive a photon index
of 1.58 $\pm$ 0.03 and a hydrogen column density of 1.1$\pm$ 0.3
$\times$10$^{22}$cm$^{-2}$ ($\chi^2_{\nu}$=1.01). The unabsorbed 2-20
keV flux with this model is 6.9$\times$10$^{-11}$ ergs cm$^{-2}$
s$^{-1}$ which corresponds to a luminosity of 2.8$\times$10$^{35}
(d/6$kpc)$^2$ ergs s$^{-1}$

\section{Discussion}

The natural interpretation of the 35 day and 103 second periods found
from our observations is that they represent the orbital period of
a neutron star in a 35 day period eccentric orbit, rotating with
a 103 second period.
We note that Reig et al. (1997) have proposed a correlation between
\halpha\ equivalent width and orbital period in Be/neutron star
binaries. This is postulated to occur because the orbit of the neutron
star acts in some way to truncate the Be star circumstellar
envelope.  The 34.5 day orbital period for \src\ derived 
from the ASM data is consistent with this
correlation (Reig, private communication). We have also examined the Ariel V SSI data presented
by Steiner et al. (1984) for any evidence of a \sqig 35 day period.
However, these data are too sparse to enable any strong
conclusions to be drawn. In addition, the source is not present in
the archival Ariel V or Vela 5B ASM light curves available at the
HEASARC at GSFC.

The parameters of \src, derived from our RXTE observations, when
plotted on a orbital period/spin period digram (cf. Corbet 1984, 1986),
show a longer pulse period (or shorter orbital period) than would be
expected from the general correlation between orbital and pulse periods
observed for Be star systems. We note the the 195.6s pulsar \2058\
which is suspected to be a Be star system also
appears to share this property. In the case of \2058\ the discrepancy
arises if
the \sqig55 day period derived from RXTE ASM observations is used (Corbet,
Peele, \& Remillard 1997). If, however, the BATSE period of twice this
value is taken (Wilson, Strohmayer, \& Chakrabarty 1996, Bildsten et al.
1997) then \2058\ {\em does} lie on the correlation curve. We note
that, in the case of the {\em supergiant} system 4U1223-624, a small
flare at {\em apastron} can sometimes also be seen (e.g. Pravdo et al.
1995) as well as a large flare near periastron. The mechanism that
produces this effect could perhaps be invoked to explain the difference
in orbital periods obtained for \2058\ from the RXTE ASM and BATSE data
sets if, for these sources, an apastron flare could be of the same
level as a periastron flare at some times. For Be/neutron star
it has sometimes been suggested that these systems might
display two outbursts per cycle if the plane of the Be star
envelope and the orbital plane are offset
(Shibazaki 1982, Priedhorsky \& Holt 1987). Such a system could
then exhibit flares when the neutron star passed through the
circumstellar envelope, in contrast to the single flare per
orbit at periastron that has so far actually been observed
for Be/neutron star binaries.
If \src\ were to display two approximately equal flares per orbit,
then its orbital period would this be \sqig 70 days and would be
in better
agreement with the overall Be star orbital/pulse period correlation.
The ASM observations do not show any compelling evidence for
an underlying \sqig70 day period. Although there is a small peak 
in the Fourier Transform at the corresponding frequency (Figure 3),
the data folded on this period (Figure 7) do not show a strong
difference between ``odd'' and ``even'' cycles of the \sqig35 day
cycle.
The simplest interpretation may be that, in fact,
\src\ does not lie on the general relationship between orbital and
pulse periods.  This correlation is believed to arise due to a
balance, on average, between spin-up and down torques (e.g. Corbet 1984,
Waters \& van Kerkwijk 1989). The equilibrium period is then dependent
on the mass transfer rate which itself should depend on the orbital period.
The strength of the correlation would thus result from the other
properties
of most of the Be star systems (e.g. Be star
circumstellar envelopes and neutron star magnetic fields)
being rather similar. \src, and
perhaps \2058, may simply have different properties from the average
system.  Indeed, Negueruela et al. (1996) propose that the spectral
type of the optical counterpart of \src\ is unusually early. This could,
for example,
perhaps result in a different circumstellar environment in \src\ compared to
other Be star systems.
For \2058\ there is no secure optical counterpart although
two candidates have been suggested by Castro-Tirado (1996).

To unequivocally determine the orbital periods of both \src\ and
\2058\ it appears that Doppler curves for these sources must be
obtained which would also give measures of system eccentricities. If
the orbital periods of these systems are indeed relatively short, then
their deviations from the overall correlations will make these systems
of interest for further study.

The spectrum that we find from the PCA is typical of X-ray pulsars. The
hydrogen column density at 1.1$\pm$ 0.3 $\times$10$^{22}$cm$^{-2}$,
while not strongly constrained, is somewhat higher than but broadly
consistent with, the estimates of 0.35$\pm$0.02 and 0.44$\pm$0.12
$\times$ 10$^{22}$ cm$^{-2}$ derived by Negueruela et al. (1996) and
Corbet \& Mason (1984) respectively from measurements of optical
reddening.

\section{Conclusion}
RXTE observations of \src\ indicate a pulse period of 103.2 seconds and a
most likely orbital period of \sqig35 days. Additional more extensive
pulse timing observations would be valuable to determine the orbital
parameters of this system.

\acknowledgments
This paper made use of quick look data provided by the RXTE ASM team at
MIT and GSFC. We thank many colleagues in the RXTE team for useful
discussions and, in particular, T. Takeshima for help with the slew
data, and A. Levine and R. Remillard  for advice on ASM data.

\pagebreak
\noindent
{\large\bf Figure Captions}

\figcaption[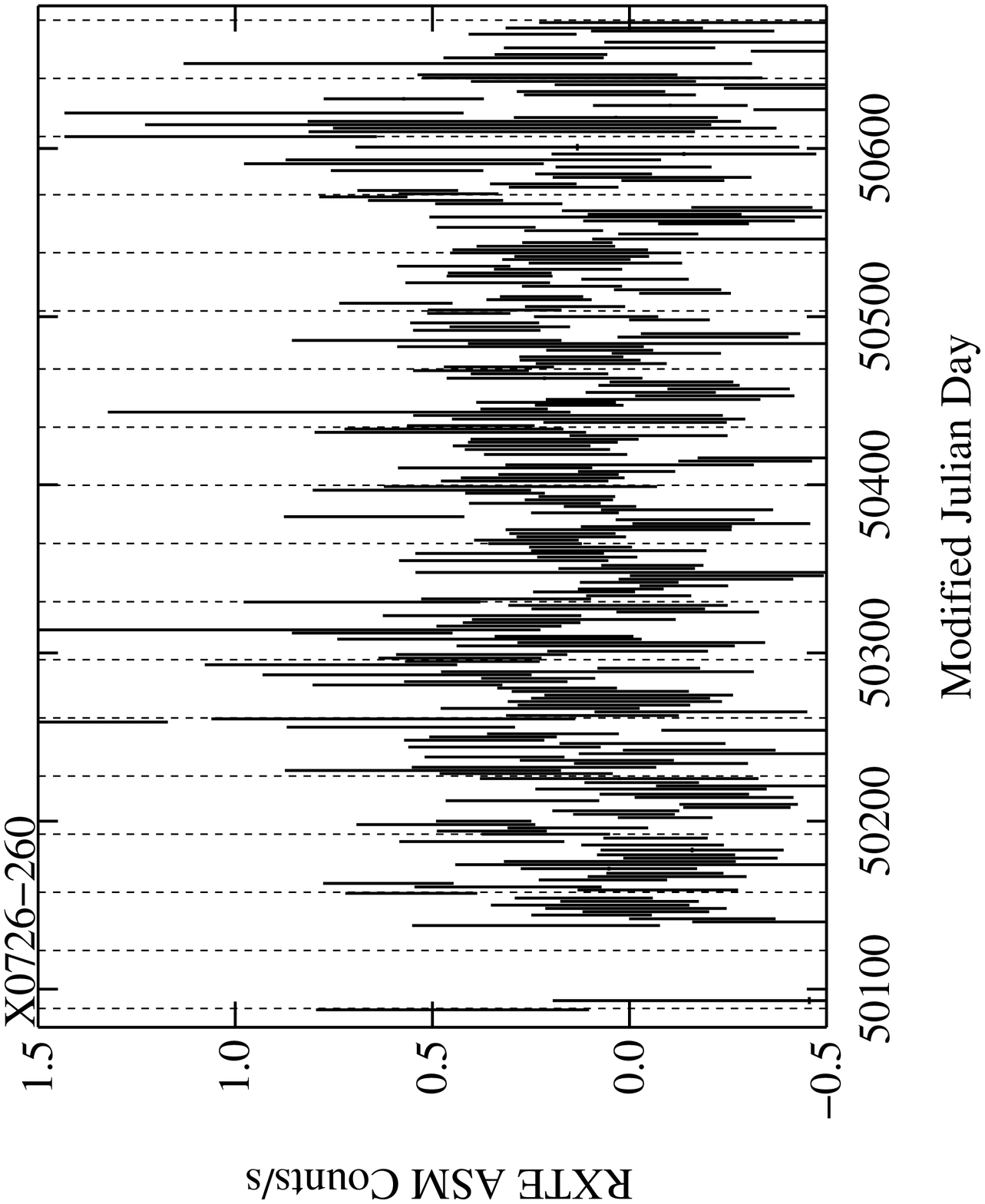]
{The light curve of \src\ from the RXTE
ASM in two day bins. The vertical dashed lines show times of expected
maximum flux based on the \sqig35 day period.}

\figcaption[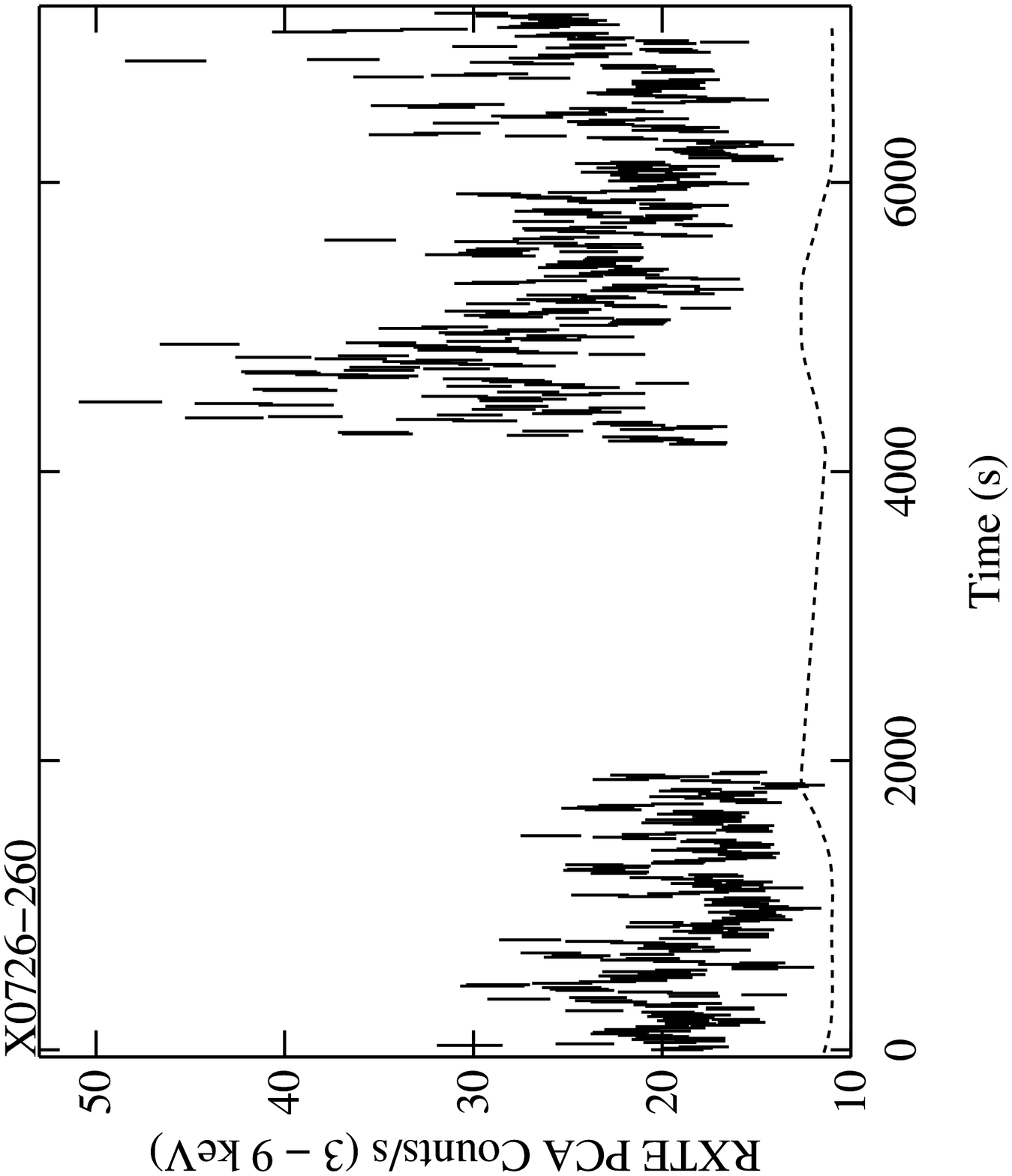]
{RXTE PCA light curve in 10 second bins for the 3 to 9 keV range. Time
is relative to 1997 June 7 02:55:28. The gap in the middle of the
observation is caused by an Earth occultation and the estimated background
level is indicated by the dashed line.}

\figcaption[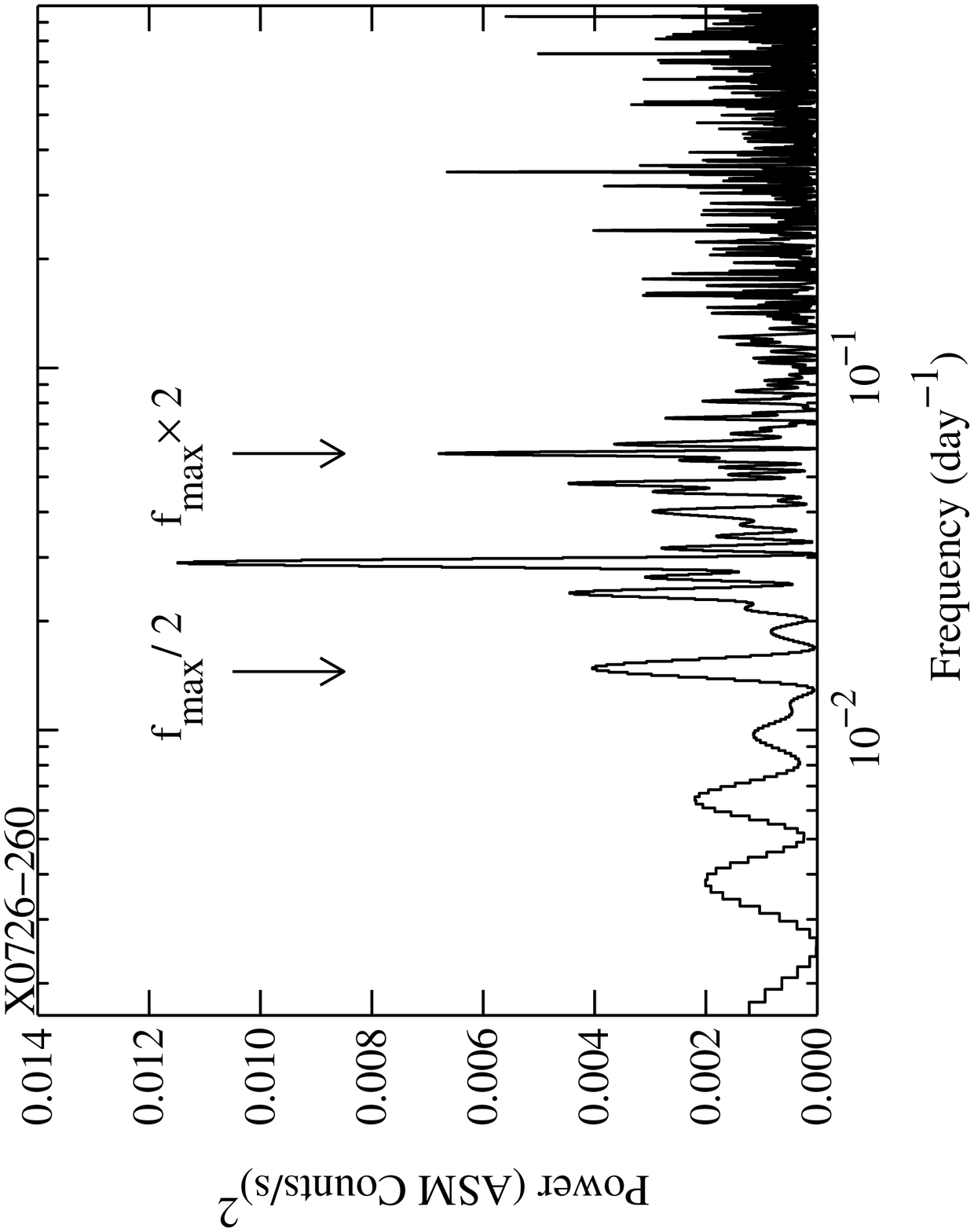]
{Fourier Transform of the RXTE ASM light curve (individual dwells).
Locations of the harmonic and sub-harmonic of the strongest peak are
indicated.  Data points were weighted by their errors in calculating
the Transform.
}

\figcaption[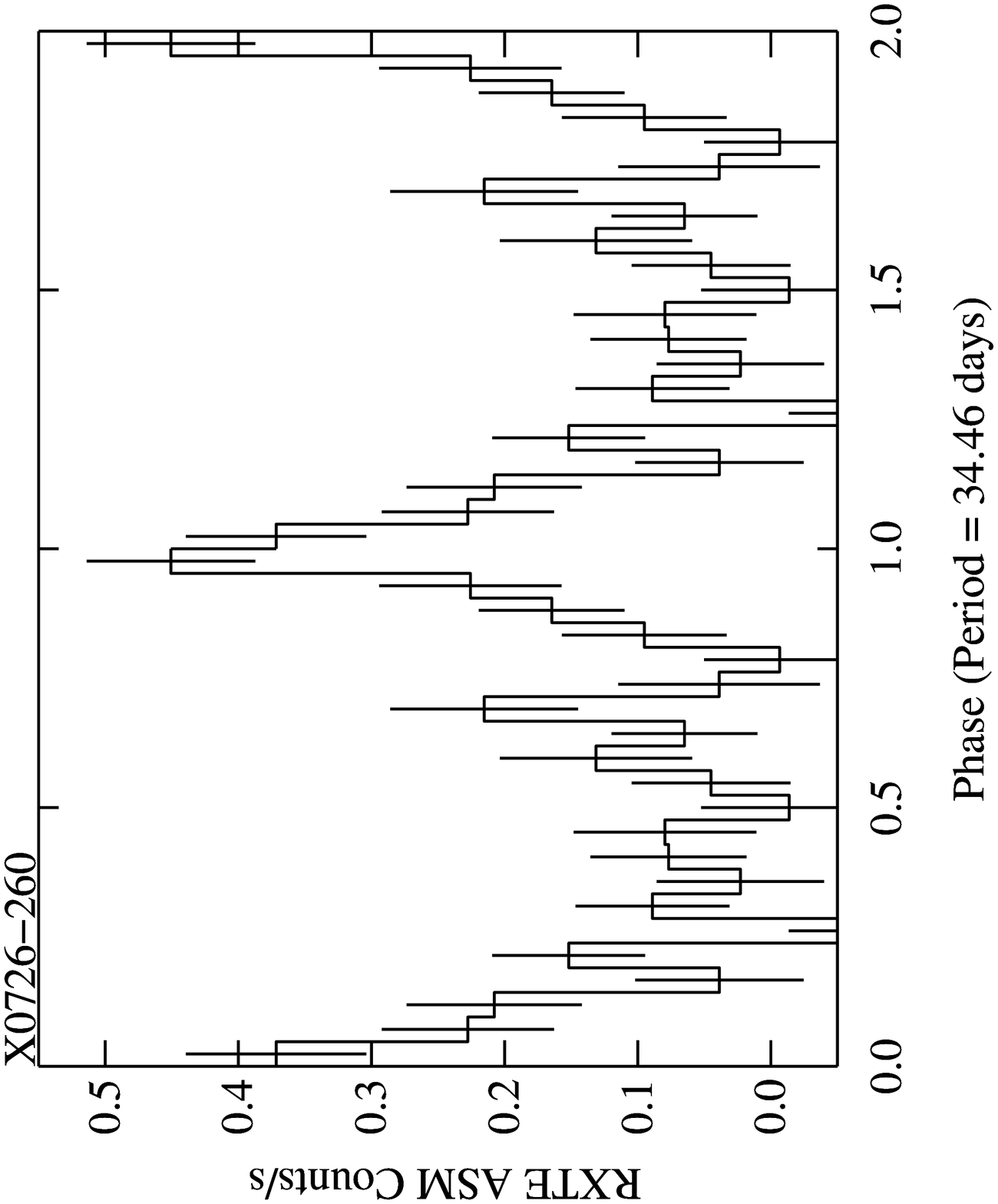]
{RXTE ASM light curve folded on the strongest peak
in the Fourier Transform. Two cycles are plotted for clarity.
}

\figcaption[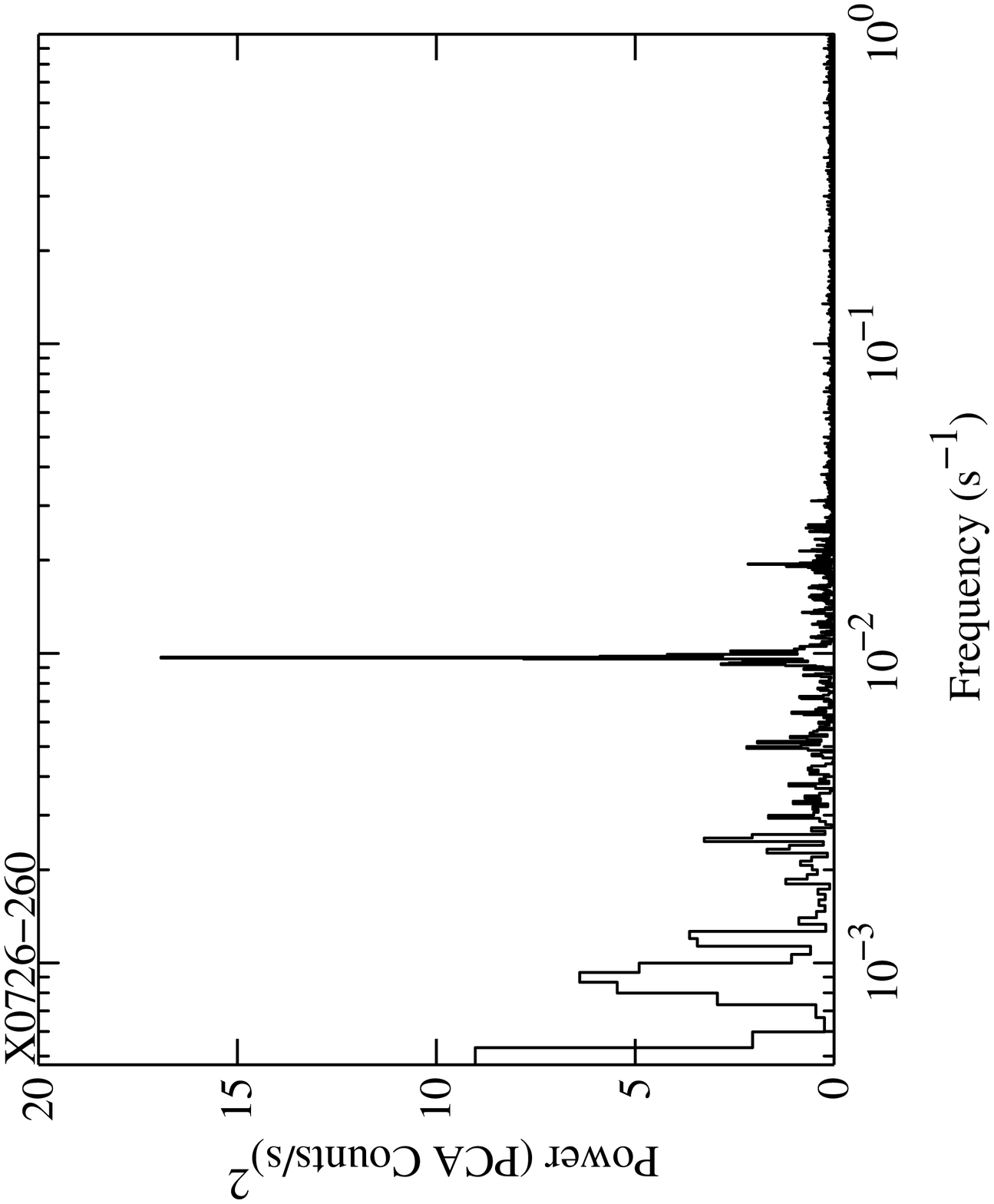]
{Fourier Transform of the RXTE PCA light curve.}

\figcaption[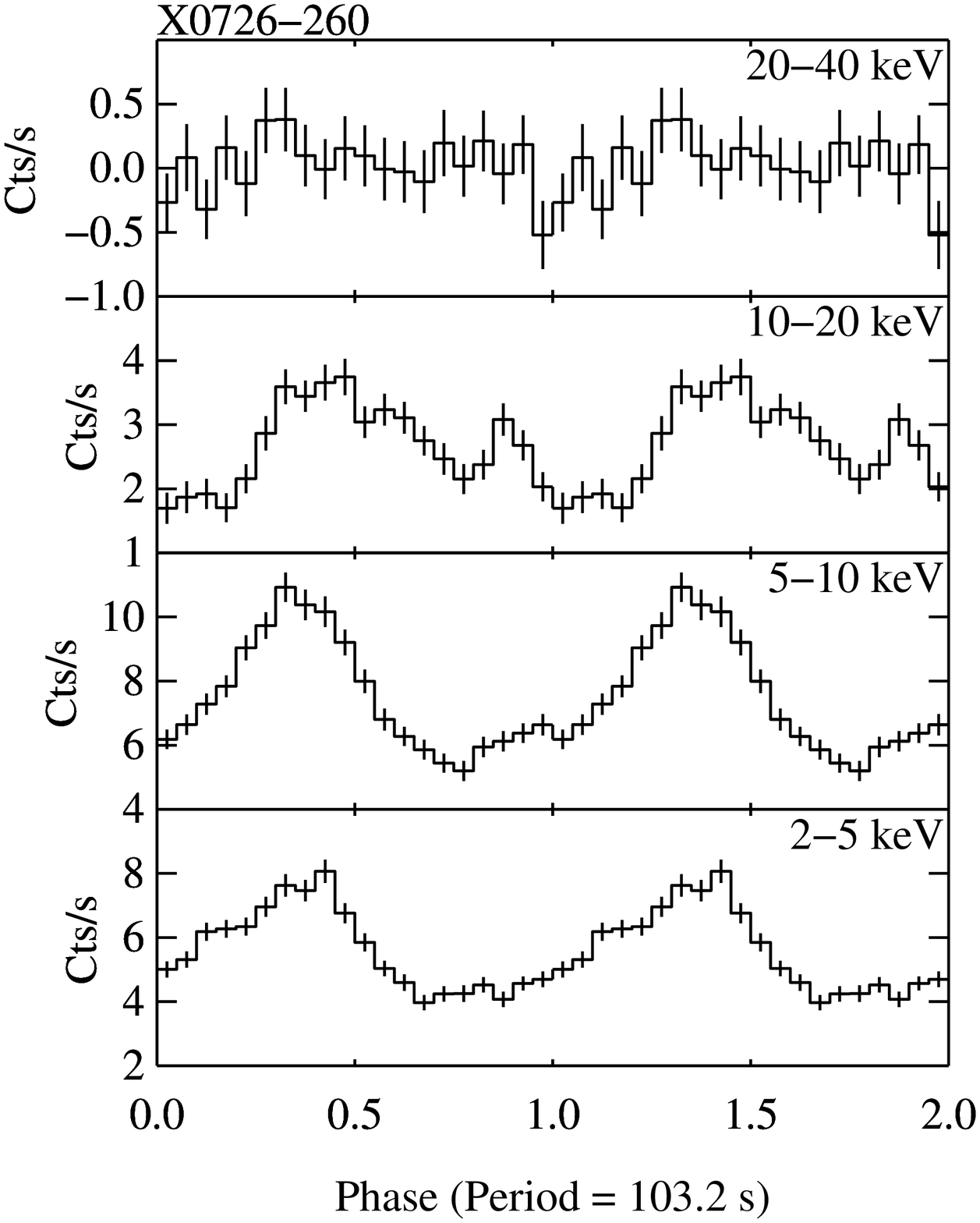]
{RXTE PCA light curve folded on the strongest peak in the Fourier
Transform as a function of energy. Estimated background levels have
been subtracted from the light curves. Two cycles are plotted for
clarity and phase is arbitrary.
}

\figcaption[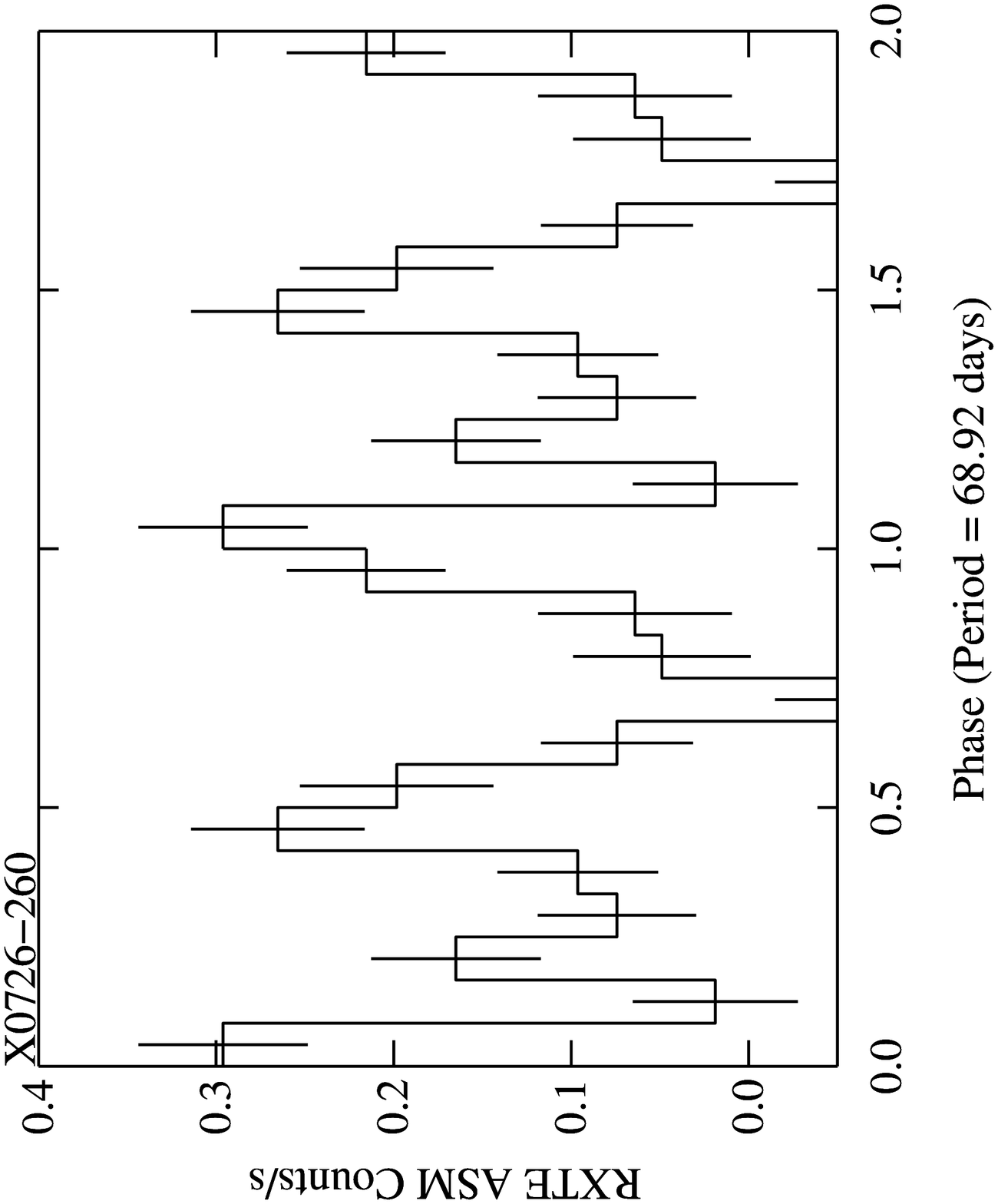]
{RXTE ASM light curve folded on a period equivalent to twice the period
of the strongest peak in the Fourier Transform. Two cycles are plotted
for clarity.
}

\begin{figure}
\plotone{figure1.ps}
\end{figure}

\begin{figure}
\plotone{figure2.ps}
\end{figure}

\begin{figure}
\plotone{figure3.ps}
\end{figure}

\begin{figure}
\plotone{figure4.ps}
\end{figure}

\begin{figure}
\plotone{figure5.ps}
\end{figure}

\begin{figure}
\plotone{figure6.ps}
\end{figure}

\begin{figure}
\plotone{figure7.ps}
\end{figure}

\end{document}